\documentclass[10pt,conference]{IEEEtran}
\IEEEoverridecommandlockouts

\usepackage{cite}
\usepackage{amsmath,amssymb,amsfonts}
\usepackage{algorithmic}
\usepackage{graphicx}
\usepackage{textcomp}
\usepackage{xcolor}
\usepackage[hyphens]{url}
\usepackage{xurl}
\usepackage{balance}
\usepackage{microtype}
\usepackage{booktabs}

\def\BibTeX{{\rm B\kern-.05em{\sc i\kern-.025em b}\kern-.08em
    T\kern-.1667em\lower.7ex\hbox{E}\kern-.125emX}}
\begin{document}
\bstctlcite{compactrefs}

\title{Deliberate Pacing for Human Oversight in Agentic~Software~Development}

\author{
\IEEEauthorblockN{Tobias Reichel}
\IEEEauthorblockA{\textit{Heidelberg University}\\
Heidelberg, Germany\\
tobias.reichel@uni-heidelberg.de}\\
\IEEEauthorblockN{Maria Teresa Baldassarre}
\IEEEauthorblockA{\textit{University of Bari}\\
Bari, Italy\\
mariateresa.baldassarre@uniba.it}
\and
\IEEEauthorblockN{Michèle Stassen}
\IEEEauthorblockA{\textit{Eindhoven University of Technology}\\
Eindhoven, Netherlands\\
m.a.stassen@student.tue.nl}\\
\IEEEauthorblockN{Sebastian Baltes}
\IEEEauthorblockA{\textit{Heidelberg University}\\
Heidelberg, Germany\\
sebastian.baltes@uni-heidelberg.de}\\
\IEEEauthorblockN{Steffen Herbold}
\IEEEauthorblockA{\textit{University of Passau}\\
Passau, Germany\\
steffen.herbold@uni-passau.de}
\and
\IEEEauthorblockN{Anamaria Mojica-Hanke}
\IEEEauthorblockA{\textit{University of Passau}\\
Passau, Germany\\
anamaria.mojica-hanke@uni-passau.de}\\
\IEEEauthorblockN{Alexander Serebrenik}
\IEEEauthorblockA{\textit{Eindhoven University of Technology} \\
Eindhoven, Netherlands\\
a.serebrenik@tue.nl}
}

\maketitle
\begin{abstract}
Agentic AI increasingly enables software development activities to proceed without continuous human involvement.
While this autonomy may accelerate development, it also raises questions about when developers can meaningfully oversee software development done by agents.
We propose deliberate pacing: intentionally adjusting when and how autonomous development proceeds to preserve opportunities for meaningful oversight.
Drawing on an exploratory multivocal literature review of 20 sources, we identify three factors motivating this vision: changing relative costs of human involvement, the timing of oversight opportunities, and interdependencies between pacing decisions across the software development life cycle.
We argue that neither uninterrupted autonomy nor frequent human intervention is universal.
Instead, pacing should balance opportunities to understand and influence consequential decisions against costs in human attention, development time, and rework.
We outline a research agenda to investigate when deliberate pacing is beneficial, how intervention timing affects oversight, and how pacing decisions interact across development activities.
\end{abstract}

\begin{IEEEkeywords}
agentic AI, human oversight, software development processes, deliberate pacing.
\end{IEEEkeywords}

\section{Introduction}

Contemporary software engineering approaches prioritize acceleration and adaptable processes, such as faster release cycles, more frequent feature releases, quicker review processes, and increased developer productivity~\cite{mantyla,10.1145/3084100.3084107,DBLP:journals/ese/KudrjavetsR24}. 
Recent industry and research narratives 
frame AI as a means of further accelerating software development \cite{Fawareh2026,10.1109/ICSE-SEIP66354.2025.00060}.

Historically, many software development activities and decisions required direct human involvement, even as automation reduced the time needed for individual steps.
AI agents can increasingly perform such activities without continuous human involvement, although their effects on development speed vary across studies.
A controlled experiment reported large speedups for an isolated programming task \cite{microsoft23peng}, while experienced open-source developers worked more slowly with AI tools on their own repositories \cite{BeckerMETR2025}.
A follow-up to the latter study found indications of a speedup instead, although the authors caution that changes in participant selection make the true effect difficult to measure~\cite{BeckerMETR2026}.
Practitioners further argue that understanding, debugging, and modifying code, rather than writing it, are important bottlenecks that faster code generation does not necessarily resolve~\cite{harsh2026blog}.
Whatever the exact productivity gains, agentic AI changes how development activities are performed by allowing agents to make and implement decisions while developers attend to other work.

This shift raises concerns about development approaches prioritizing speed.
Agentic AI enables substantial parts of a software system to be prototyped with limited human involvement, potentially allowing consequential decisions to be implemented before developers evaluate them.
Studies report risks associated with AI-assisted development, including delivery stability~\cite{dora2025} and security~\cite{FU2025,Perry2023}.
These findings point to the importance of activities beyond code generation, such as design, problem analysis, compliance assurance, and critical evaluation of AI-generated artifacts~\cite{liang2024large,lee2025,sakhalkar2025,barenkamp2020}.
Studies of AI's impact on the software development life cycle (SDLC) likewise emphasize continued human involvement in planning and testing~\cite{gonzalezrueda2026}.

However, human involvement does not necessarily ensure meaningful oversight. 
Developers must have sufficient information to understand relevant decisions and opportunities to influence them.
Consider an agent that selects an architectural dependency and proceeds to implement the corresponding components and tests without requesting human input.
By the time a developer reviews the results, several artifacts may depend on that decision, and reconsidering it may require substantial rework.
Intervening earlier could avoid such costs, but may also interrupt productive autonomous execution before sufficient information is available for an informed decision.
Deciding where developers should intervene is therefore not enough. The timing also determines whether their involvement can influence autonomous development.

In this paper, we distinguish between \emph{speed} and \emph{pace}. 
\emph{Speed} is how quickly software is developed, measured as output per unit of time. 
\emph{Pace} is how that speed is distributed over time. 
We define \emph{deliberate pacing} as intentionally adjusting when and how autonomous development proceeds to create opportunities for meaningful human oversight.
Such adjustments may include pausing an agent before consequential decisions are implemented, postponing subsequent activities until a human review is completed, or changing when intermediate results are presented for evaluation.
Not all oversight requires slowing down autonomous execution. Developers may also monitor ongoing activities or review completed work asynchronously.
Whether deliberate pacing is beneficial therefore depends on the opportunities for oversight that creates relative to its costs in time, human attention, and development throughput.

We present \textbf{a vision of deliberate pacing as a means of preserving meaningful human oversight in agentic software development}.
Our central proposition is that the timing of human intervention matters: retrospective review may not provide the same opportunities to influence autonomous development as intervention before consequential decisions are implemented.
Drawing on a multivocal literature review (MLR) of 20 sources on slowing down software development, we examine three factors that motivate this vision: the changing relative costs of human involvement, the timing of opportunities for meaningful oversight, and the interdependencies between pacing decisions across the SDLC.
Rather than proposing universal checkpoints or advocating slower development as an end in itself, we outline a research agenda for investigating when, where, and how deliberate pacing enables meaningful oversight more effectively than uninterrupted autonomous execution followed by retrospective evaluation.
\section{Vision}
\label{sec:vision}

We argue that deliberate pacing preserves opportunities for meaningful human oversight by determining when agents act autonomously and when human intervention is needed. Prior research shows that developers' autonomy preferences vary with experience, motivating configurable autonomy~\cite{Ghorbani2023Autonomy}. Agentic AI adds a temporal dimension: not only how much autonomy to grant, but when to interrupt it for human oversight. Three factors motivate our vision.

\textit{First, agentic AI changes the relative costs and opportunities for human involvement.}
Traditionally, software development required substantial human effort to produce artifacts, even when developers relied on automated tools. Agentic AI can perform increasingly complex development activities with limited human intervention, potentially reducing execution time considerably. Consequently, activities such as reviewing designs, validating requirements, and understanding generated code may account for a growing proportion of development effort. This does not necessarily make human involvement more expensive in absolute terms, but changes its relative cost compared to autonomous execution. Although developers have long performed other work while automated tools execute, agentic AI extends asynchronous execution to activities involving design and implementation decisions. 
Earlier research on code review bots already illustrates that automating development activities can affect both throughput and developer communication~\cite{Wessel2022Quality}.
Deliberate pacing decisions must therefore consider not only how much an intervention delays autonomous execution, but also how much human attention it requires and what opportunities for oversight it creates.

\textit{Second, the timing of human involvement shapes opportunities for meaningful oversight.}
Developers have long delegated work to automated tools, but increasingly autonomous agents can also make decisions about requirements, design, implementation, and testing. Human involvement may consequently shift from performing these activities to supervising, validating, or revising decisions made by agents. However, involving a human does not necessarily ensure meaningful oversight: developers must have sufficient information to understand relevant decisions and the opportunity to influence them. In particular, reviewing an agent's work after completion is not always equivalent to intervening before consequential decisions are implemented.
Conversely, intervening too early may provide insufficient information for an informed decision. Deliberate pacing can create opportunities for oversight at moments when developers have both sufficient information and the ability to act upon it. Without such opportunities, teams may also accumulate cognitive and intent debt~\cite{storeyTechnicalDebtCognitive2026}, as developers lose shared understanding of the system or decisions diverge from the goals and constraints intended to guide development.

\textit{Third, pacing decisions must be considered across the entire SDLC.}
Meaningful oversight may be enabled through different interventions at different stages of development, and their costs and benefits cannot always be assessed independently. For example, human validation of requirements may prevent misunderstandings from propagating into design and implementation, while independent validation of test oracles may identify discrepancies between implemented behavior and intended functionality. However, these interventions are not necessarily interchangeable: Reviewing an implementation cannot always compensate for requirements that fail to reflect stakeholder needs. Furthermore, delaying an intervention may allow autonomous agents to build subsequent decisions upon assumptions that have not yet been validated, potentially increasing the cost of later corrections. Conversely, frequent interruptions may impose unnecessary delays and coordination overhead without improving outcomes. 
We therefore argue that pacing decisions should be evaluated not only by their immediate costs and benefits, but also by how they affect prior and subsequent opportunities for meaningful oversight across the SDLC.

Together, these factors motivate our vision of  \textit{deliberate pacing as a means of enabling meaningful human oversight without unnecessarily sacrificing the benefits of autonomous development}. 
Deliberate pacing may involve pausing an agent before consequential decisions are implemented or postponing subsequent activities until human evaluation is complete. However, meaningful oversight may also be achieved without slowing down autonomous execution, for example through asynchronous monitoring or retrospective review. 
The challenge is to determine when deliberate pacing offers benefits that cannot be achieved as effectively through uninterrupted autonomous execution.
\section{Foundations for Deliberate Pacing}

To inform our vision, we conducted an exploratory MLR of 20 sources (8 academic and 12 grey literature), following the guidelines of Garousi et al.~\cite{GAROUSI2019101}.
Table \ref{tab:sq} lists the search terms we used in ACM Digital Library, IEEE Xplore, and Scopus. For the grey literature, we queried Google Search with ``slow software development", ``slow programming movement", and ``slow code movement". We ran the search in an incognito browser window while logged out of Google to minimize personalization. We filtered the search results based on whether they helped answer our question: ``Why and under which circumstances do individuals, teams, and organizations intentionally slow down in software development?''

Through inductive thematic analysis, we identified five recurring themes: slowing down for software quality, slowing down for long-term sustainability, slowing down for care and ethics, slowing down as resistance to acceleration, and developing at an appropriate pace. The first three concern motivations for slowing down, while the last two address the broader questions of why speed is prioritized and how an appropriate pace can be determined.
\begin{table}[htbp]
  \centering
  \caption{Search Query for Bibliographic Databases}
  \label{tab:sq}
  \begin{tabular}{@{}p{\linewidth}@{}}
  \toprule
    (``software development" OR ``software engineering")
     AND \\ ( (``slow food" OR ``slow tourism" OR ``slow travel" OR ``citta slow" OR ``cittaslow" OR ``slow city" OR ``slow housing" OR ``slow living" OR ``slow art" OR ``slow cinema" OR ``slow design" OR ``slow journalism" OR ``slow education" OR ``slow philosophy" OR ``slow writing" OR ``slow reading" OR ``slow science" OR ``slow research" OR ``slow university" OR ``slow work" OR ``slow management" OR ``slow fashion" OR ``slow consumption" OR ``slow technology")
     OR \\ (``mindful development" OR ``mindful software" OR ``deliberate development" OR ``deliberate software" OR ``reflective software" OR ``reflective development" OR ``thoughtful software" OR ``reflective development") AND (``speed" OR ``slow" OR ``fast" OR ``rhythm" OR ``pace" OR ``temporal")
     OR \\ (``good enough software" OR ``good enough development") )\\
 \bottomrule
  \end{tabular}
\end{table}

Slower and more thoughtful software development, which is characterized by an emphasis on careful design and writing quality code, can contribute to software quality~\cite{ventrella}. This emphasis does not necessarily imply that extensive design up front is needed~\cite{nelson}. Code can be rewritten and redesigned, as long as a developer pauses to think and takes the time to create something worth making~\cite{nelson}. An important part of carefully thinking through solutions is reducing complexity by asking whether a feature or piece of code is worth adding~\cite{sean_no, david, zechner, ascetic}.
Several grey literature sources in our MLR link slowing down to software craftsmanship. They argue that software craftsmanship requires careful thinking, focus, deliberate practice, and a deep understanding of the intended result~\cite{craft_slow, ascetic, frank}. None of the academic sources in our MLR mention slowing down as a core aspect of software craftsmanship, although it has been studied in academic work outside our sample, for example in a literature review by Sundelin et al.~\cite{anatomy_soft_craft}.
However, two sources note that rapid development is sometimes necessary or desirable~\cite{developing_future, mellis_1998}. In rapidly changing environments such as dynamic markets, it is essential for a company's success to respond quickly to improve its competitiveness~\cite{developing_future, mellis_1998}. The goal shifts from delivering quality to delivering sufficient quality on time to gain or keep a competitive advantage~\cite{mellis_1998}, but as a company grows and the product evolves, the quality should be incrementally increased~\cite{developing_future}. A quickly developed minimum viable product can be used to test the market and confirm that there is demand. Afterwards, developers must take time to code with quality and maintenance in mind to minimize the previously accrued technical debt and build a more reliable product~\cite{shaw}.

A second recurring rationale for slowing down software development is the long-term sustainability of software systems as well as software development practices. The underlying problem is an imbalance in how innovation and maintenance are valued~\cite{timeless_programming, resource_scarcity}. An uncritical focus on innovation overlooks what happens after innovation, and how much everyday society depends on stable, older technologies~\cite{resource_scarcity}. Due to the fast pace of innovation, developing with maintenance and longevity in mind requires deliberately taking time. Another aspect of slowing down for sustainability is critically evaluating a project's dependencies so that fewer dependencies need to be kept up to date and unnecessary complexity is reduced~\cite{timeless_programming, david}. Crichton discusses the need for a ``slow code'' movement toward stable libraries and programming languages that preserve backward compatibility~\cite{craft_prog}. 

The theme of slowing down for care and ethics comes mainly from Human-Computer Interaction (HCI) research, which has argued for slowing down design processes to create space for critical reflection, ethical deliberation, and community involvement \cite{temporal, temporal_awareness}. For example, a case study of community-centered design describes tensions between responding quickly to urgent crises and taking time to build trust with community partners \cite{temporal}. While rapid action may be necessary in urgent situations, excessively prioritizing speed can limit opportunities for reflection and sideline community perspectives \cite{temporal, temporal_awareness}. In this context, slowing down means creating time for reflection, discussion, and engagement with those affected by the system. Lack of time within a project can lead to technical elements being prioritized over ethical deliberation and can conflict with the time required for participatory community research~\cite{temporal_awareness}. Rapid iteration and market-driven pressures can further hinder ethical considerations \cite{who_speaks}. Meanwhile, prioritizing high-quality standards over development speed can improve a development team's ability to identify and address ethical concerns \cite{who_speaks}. Software security shows similar tensions~\cite{compl_sec, TONDEL2022102744}. A project's ability to maneuver depends on the available time, budget and competence, as well as the type of software project and stage of development. 
Under tight time constraints, security may be deprioritized, especially when security is viewed as a cost without real gain~\cite{TONDEL2022102744}.

Sources in the theme of slowing down as resistance to acceleration question the assumption that delivery speed, developer productivity, and innovation are inherently desirable and asking why and for whom software is built. A recurring critique in the sources is that of Silicon Valley's development culture, particularly Facebook's slogan ``move fast and break things''~\cite{reuters_zuckerberg_letter_2012}. Halperin and McElroy~\cite{temporal} criticize the temporal logic of acceleration and instantaneity, which views time as a quantitative resource to optimize. Disruptive companies set the ideals against which software development is measured. However, in what Bialski calls ``middle tech'', where software is more often maintained rather than developed as a disruptive innovation, a negotiated culture of ``good enoughness'' reconciles ideals of innovation with everyday compromise~\cite{Bialski_2024}. The sources also challenge growth-driven economic systems. Developers may not be able to restructure these systems, but they can change the processes that guide development~\cite{who_speaks}, for example by incrementally improving code quality in a start-up~\cite{developing_future}, designing more slowly~\cite{ventrella}, or creating room for discussion and critical reflection~\cite{who_speaks}. Pargman and Wallsten~\cite{resource_scarcity} calls for questioning who benefits from new technologies and who is disadvantaged by them.

The final theme, developing at an appropriate pace, addresses how teams meet the need to stay competitive and respond to urgent situations while still taking the time that quality, maintainability, ethical deliberation, and stakeholder engagement require. Developing at an appropriate pace means spending the time that is needed on the relevant dimensions given the context, problem, stakeholders, and aims of the project. What constitutes `sufficient' time cannot be determined universally ahead of development as it is a situational decision that requires continuous negotiation~\cite{Bialski_2024}. Developing at an appropriate pace therefore involves moving between varying development speeds and balancing between care and compromise~\cite{Bialski_2024}, urgency and care \cite{temporal}, and innovation and maintenance~\cite{timeless_programming, resource_scarcity}.
Except from the sources that discuss AI coding, such as Zechner~\cite{zechner}, the sources in our MLR assume that developers write the code themselves, so the time they take for reflection is also the time they spend producing the code. Agentic AI separates the two. An agent can implement a feature, add a dependency, or refactor a module quickly, while the reflection that our identified themes propose still takes human time. The questions that the sources raise remain relevant, such as whether a feature~\cite{sean_no, zechner, ascetic} or dependency is worth adding~\cite{timeless_programming, david, craft_prog} or whose perspectives a design takes into account~\cite{temporal_awareness, temporal}. With agents, however, developers need to ask these questions at points in the SDLC where they can still act on the answer. Developing at an appropriate pace then means deciding when an agent proceeds on its own and when it waits for human judgment.
\section{Future Plans}

To develop our vision into a full-length research
contribution, we propose three research questions
concerning deliberate pacing and meaningful human
oversight in agentic software development.

\textit{RQ1: When does meaningful human
    oversight benefit from deliberate pacing?}
Through developer interviews, recorded development
sessions, and agent execution traces, we will examine
when developers allow agents to proceed uninterrupted and when they introduce deliberate pauses for human evaluation. We will identify consequential decisions and assess whether interventions enable developers to recognize potential problems, question assumptions, and influence subsequent agent actions. We will also examine the costs of these interventions in human attention, development time, and avoided or incurred rework. These observations will help identify circumstances in which deliberate pacing offers benefits that justify its costs.

  \textit{RQ2: How does the timing of pacing
    interventions affect meaningful human oversight?}
We will compare different arrangements of human
oversight, including continuous monitoring,
deliberate pauses before consequential decisions
are implemented, and retrospective review~\cite{Dhanorkar2026Human}.
We will assess developers' ability to identify,
understand, and influence agent decisions, the emotions 
they experience~\cite{Grassi2026Selfmonitoring} 
and express~\cite{Cassee2025Negativity} during oversight, as
well as the associated costs in human effort,
development time, and subsequent rework.

\textit{RQ3: How do pacing interventions
    across the SDLC interact in enabling meaningful
    human oversight?}
We will investigate how interventions in one development activity affect the need for and effectiveness of oversight in subsequent activities, including their combined costs and benefits.
Since controlled experiments would require us to study different interventions and their interactions in a longitudinal manner across the whole SDLC, they are very high-effort and, therefore, only suitable to test approaches for which there is already evidence for their effectiveness. To address this, we will develop new research methods.
On the one hand, we will explore quasi-experiments~\cite{Kampenes2009} as substitue for randomized control trials. Such experiments can combine data from software repositories with developer surveys to explain the underlying processes.
On the other hand, simulation-based methods~\cite{herbold2026softwareteamspendagents} will estimate how different pacing approaches interact.

Together, these studies will provide empirical evidence on when deliberate pacing is warranted and inform principles for coordinating autonomous execution and meaningful human oversight across the SDLC.
\section{Conclusion}

In this paper, we present a vision of deliberate pacing as a means of preserving meaningful human oversight in agentic software development. We motivate our vision by a MLR on slow software development. We identified five recurring themes in 20 sources (8 academic and 12 grey literature) such as slowing down for software quality, for long-term sustainability, care and ethics, as resistance to acceleration, and developing at an appropriate pace. We argue that agentic AI changes the conditions under which speed, quality and human involvement must be balanced. Deliberate pacing should be assessed by how much an intervention delays the execution, how much human attention it requires, and what opportunities for human oversight it creates. Oversight requires that developers have sufficient information and can act on it at the moment of intervention. Finally, we argue that deliberate pacing decisions must be considered across the entire SDLC. A late intervention can cost more than an earlier intervention would have.

In the future, we plan to research when human oversight benefits from deliberate pacing and how it is affected by the timing of the interventions. Additionally, we will investigate the impact of deliberate pacing across the SDLC by using new research methods which allow us to simulate these complex interactions between activities in the SDLC.

The research community must not only focus on developing such slow and fast practices, but also live by them, since we shape future generations both through our research and through our roles in the academic education system. In the past, we have often made the mistake of not following the practices that we preach: we write about best practices without following them~\cite{Mojica2026} and advocate for open science~\cite{boisvert2016incentivizing, mendez2020open}, yet often do not make our resources available~\cite{heumuller2020publish}. We should not repeat this mistake with the speed that AI seemingly offers.

\section*{Acknowledgment}
Some of the ideas presented in this paper emerged from discussions during the Shonan Meeting No. 228, ``Envisioning the Future of (eXtended Reality) Software Engineering.'' We are grateful to the meeting organizers and participants for their insightful contributions and the fruitful discussions that helped shape this work.


\balance
\bibliographystyle{IEEEtranS}
\bibliography{references}

\end{document}